%% file: Slepton_vbfnlo.tex
\documentstyle[12pt,axodraw]{article}
\def\r {\rightarrow}

\def\neu {\tilde\chi_1^0}
\def\l {\ell}
\def\slep {\tilde{\ell}}
\def\sll {\tilde{\l}_L}
\def\slr {\tilde{\l}_R}
\def\msl {m_{\slep}}

\def\beq{\begin{equation}}
\def\eeq{\end{equation}}

\def\GeV{\;{\rm GeV}}

\newcommand{\lsim}
{{\;\raise0.3ex\hbox{$<$\kern-0.75em\raise-1.1ex\hbox{$\sim$}}\;}}
\newcommand{\gsim}
{{\;\raise0.3ex\hbox{$>$\kern-0.75em\raise-1.1ex\hbox{$\sim$}}\;}}

\begin{document}

\begin{flushright}
\texttt{hep-ph/0612119}\\
\texttt{KA-TP-12-2006}\\[3ex]
\end{flushright}

\begin{center}
{\Large\bf Next-to-leading order QCD corrections to slepton pair production via vector-boson fusion} \\[10mm]
{\bf Partha Konar}
and {\bf Dieter Zeppenfeld}  \\[2ex]
{\em Institut f\"ur Theoretische Physik, Universit\"at Karlsruhe,\\ 
P.O.Box 6980, 76128 Karlsruhe, Germany}\\ 
\end{center}
\vskip.5in

\begin{abstract}

Slepton pairs can be produced in vector-boson fusion processes at
hadron colliders. The next-to-leading order QCD corrections to the 
electroweak production cross section for $pp\r\slep^+\slep^- +2$~jets  
at order $\alpha_s\alpha^4$ have been calculated and implemented
in a NLO parton-level Monte Carlo program. Numerical results are
presented for the CERN Large Hadron Collider

\end{abstract}

\vskip 1 cm
\noindent
\texttt{{\bf \sc PACS Nos:} 11.10.Gh, 12.60.Jv, 14.80.Ly} \\

\noindent
\texttt{{\bf \sc Key Words}: QCD corrections, Beyond Standard Model, Supersymmetry Phenomenology}\\

\vfill

\newpage

\noindent
{\underline {\em Introduction}}:~

Among the primary goals of the CERN Large Hadron Collider (LHC) are the
discovery of the Higgs boson, thus shedding light on the yet unexplored 
mechanism of
electroweak (EW) symmetry breaking, and the search for physics beyond
the standard model. Within the area of Higgs boson studies, 
vector-boson fusion (VBF) processes have emerged as being highly 
promising for 
revealing information on the symmetry breaking sector~\cite{vbfhiggs}. 
The prototypical
process is $qq\to qqH$, which proceeds via $t$-channel $W$ or $Z$
exchange. The two scattered quarks emerge as forward and backward
jets (called tagging jets) which provide a characteristic signature for 
VBF and allow to significantly suppress backgrounds. As a result, VBF
searches are expected to lead to quite clean Higgs boson signals.

A natural question is whether vector boson fusion is a useful tool also 
for the study of other signals of new physics. Some recent work has 
indicated the effectiveness of VBF
channels in the context of new physics searches, particularly for new
particles that do not interact strongly. Perhaps the best 
example~\cite{konar_nuchar} is
afforded by supersymmetric theories, wherein conventional search
strategies for neutralinos and charginos may run into
difficulty at the LHC, for a significant part of the parameter space.
The possibility of a slepton search  has been
studied for vector-boson fusion as well~\cite{konar_slep}. 
A more recent study~\cite{rain} 
on VBF slepton production using Smadgraph arrived at a substantially 
smaller 
cross-section, however, which is partly caused by large cancellations among
VBF-type diagrams and bremsstrahlung diagrams at the Born level.

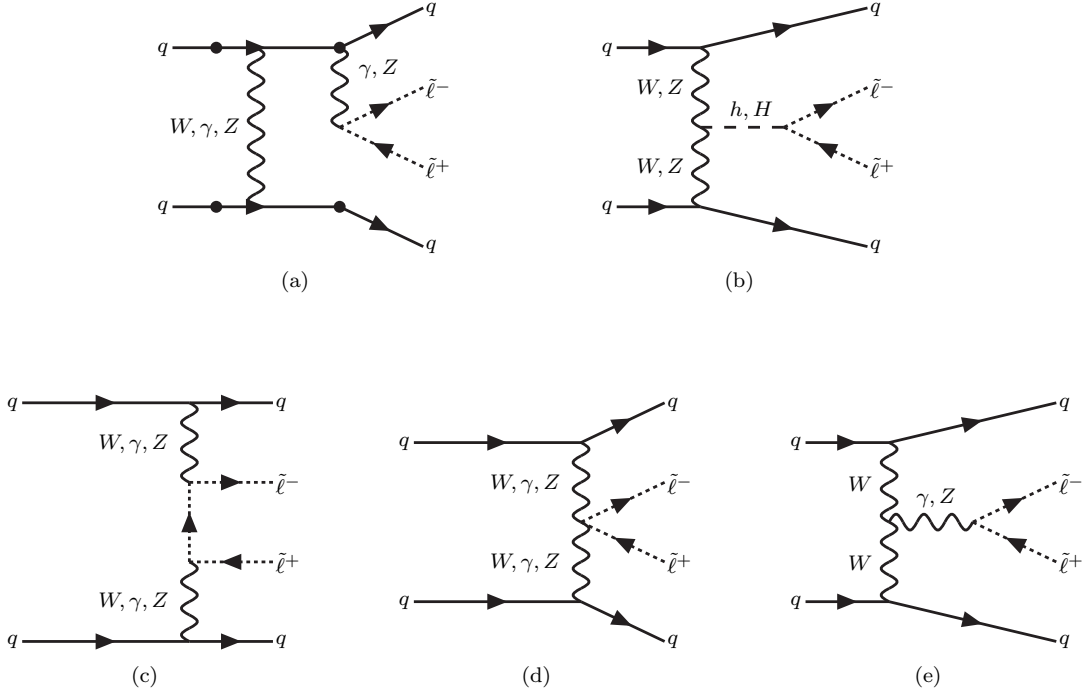
\begin{figure}[t] 
\input{diag_generic.tex} 
\caption{\em Generic LO parton level diagrams leading to slepton
  pair-production through electroweak VBF at hadron colliders.
}
\label{fig:feyn}
\end{figure}

The discrepancies between these previous results lead us to a 
recalculation of the slepton pair-production cross section in VBF. 
The relevant Feynman graphs for this process are depicted in 
Fig.~\ref{fig:feyn} for the tree level contributions. In this 
approximation, we confirm the new results of Ref.~\cite{rain}.
In addition, we also perform a calculation of the NLO QCD corrections
to this VBF process. The NLO calculation closely follows previous 
calculations for $Hjj$ and $Zjj$ production in VBF in 
Refs.~\cite{dieter_h,dieter_z}. It uses the Catani-Seymour subtraction 
scheme~\cite{Catani} for implementing the real and virtual NLO contributions
in the form of a fully flexible parton level
Monte Carlo program.

\noindent
{\underline{\em Calculation}}:~

The Feynman graphs contributing to $pp \r \slep^+ \slep^- + 2$~jets 
at tree level are indicated in Fig.~\ref{fig:feyn}. 
 Considering the possible choices of external quarks or anti-quarks, 
the sub-processes can be grouped into neutral-current (NC) processes, 
like $u c \r u c \slep^+ \slep^-$, and charged-current 
(CC) processes, like $u s \r d c \slep^+ \slep^-$.
For the purpose of calculating the 
virtual QCD corrections, the Feynman graphs are divided into 
Compton scattering type graphs, as in Fig.~\ref{fig:feyn}(a),
and the VBF type graphs as in Fig.~\ref{fig:feyn}(b-e).
The first class (Fig.~\ref{fig:feyn}(a) and three additional bremsstrahlung
diagrams, with the vector boson radiated at the position of the blobs) 
corresponds to the emission of the external vector boson from one 
of the two quark lines. The VBF graphs represent $V V \r \slep^+
\slep^-$. Here, V stands for a $t$-channel $\gamma$, $Z$ or $W$ boson.
For selectron or smuon production one expects a negligible contribution
from Fig.~\ref{fig:feyn}(b). We do include this Higgs exchange contribution
for stau pair production, however, anticipating strong enhancements of the 
couplings to the Higgs bosons at large $\tan\beta$.

Contributions from anti-quark initiated $t$-channel processes such as
$\bar{u} c \r \bar{u} c \slep^+ \slep^-$, which emerge from crossing the
above processes, are fully taken into account. 
On the other hand, two additional classes of diagrams which can appear in
case of identical quark flavors, are simplified in our calculation.
The first concern $s$-channel exchange diagrams, where
both virtual vector bosons are time-like. These diagrams correspond to 
vector boson pair production with subsequent decay of one neutral vector
boson to $\slep^+\slep^-$ while the other one decays into a
quark-antiquark pair. These contributions can be safely neglected in the
phase-space region where VBF can be observed experimentally, with
widely-separated quark jets of large invariant mass. The second class
corresponds to $u$-channel exchange diagrams which are obtained by the
interchange of identical final state 
(anti)quarks. Their interference with the $t$-channel diagrams is strongly
suppressed for typical VBF cuts and therefore neglected in
our calculation. Color suppression further reduces any interference terms.

Throughout our calculation, fermion masses are set to zero and external
b- and t-quark contributions are neglected. For the
Cabibbo-Kobayashi-Maskawa matrix $V_{CKM}$, we use a diagonal form
equal to the identity matrix. This yields the same results as a
calculation using the exact $V_{CKM}$ when the summation over all
quark flavors is performed.

The computation of NLO corrections is
performed in complete analogy to Ref.~\cite{dieter_z}.
For the real-emission contributions, we consider the diagrams with a
final-state gluon by attaching the gluon to the quark lines in all
possible ways. As a result one obtains two distinct, non-interfering 
color structures which correspond to gluon emission off the upper or off
the lower quark line in Fig.~\ref{fig:feyn}. Subprocesses with an initial
gluon are obtained by crossing the final state gluon on a given quark 
line with the incident quark or anti-quark of this same quark line. 
As a result only one color structure exists for initial gluons. The
other color structure would correspond to an $s$-channel process of the
type $gq\to VVq$, which has been neglected also at Born level.

All amplitudes are evaluated numerically using the amplitude
techniques of Ref.~\cite{dieter_helicity}. The calculation is simplified
by introducing the leptonic tensors 
$\Gamma^\alpha_V$ and $L^{\alpha \beta}_{VV}$, which describe the
effective polarization vector of the final state decay
$V(q)\to\slep^-(p_1)\slep^+(p_2)$,
\beq
\Gamma^\alpha_V(p_1,p_2) = \frac{g_\tau^{V\slep}}
{(p_1+p_2)^2-m^2_V+i m_V\Gamma_V} (p_1-p_2)^\alpha \;,
\label{eq:currect}
\eeq
and the effective sub-amplitude for the process $V_1^\alpha V_2^\beta
\to \slep^+\slep^-$. The leptonic tensor $\Gamma^\alpha_V$ is common 
to real emission graphs and Born graphs appearing in the Catani-Seymour 
subtraction terms and needs to be calculated only once at a 
given phase space point, independent of the crossing of the colored 
partons. Similarly, $L^{\alpha \beta}_{VV}$ is
only needed for two distinct momentum flows (gluon attached to the upper
or to the lower quark line) at any phase space point. It is calculated
in the complex-mass scheme~\cite{Denner:1999gp} which implements the 
Breit-Wigner propagators of the resonant $Z$-boson in a gauge 
invariant way.

At NLO, we have to deal with singularities in the soft and collinear
regions of phase space which are regularized in the dimensional-reduction
scheme \cite{Siegel} with space-time dimension $d = 4 - 2 \epsilon$.  The
cancellation of these divergences with the respective poles from the
virtual contributions is performed by introducing the counter terms of
the dipole subtraction method \cite{Catani}.  
Since these divergences only depend on the color structure of the
external partons, the analytical form of subtraction terms and finite
collinear pieces encountered for VBF slepton pair production, in terms 
of the respective Born amplitude, is
identical to the ones given in Ref.~\cite{dieter_h}.  

The virtual corrections to the amplitudes arise from a virtual gluon
emitted and re-absorbed by either the upper fermion line or by the lower
fermion line. For both contributions the resulting virtual amplitude, 
${\cal M}_V$, can be expressed in term of a divergent part, which is 
fully factorisable in terms of the original Born 
amplitude, ${\cal M}_B$, and a finite part, $\tilde{{\cal M}_V}$,
\beq
{\cal M}_V = {\cal M}_B \frac{\alpha_s(\mu_R)}{4\pi}C_F\left(
  \frac{4\pi \mu_R^2}{Q^2}\right)^\epsilon \Gamma (1+\epsilon) 
  \left( -\frac{2}{\epsilon^2}-\frac{3}{\epsilon}+c_{virt}\right)
  + \tilde{{\cal M}_V} \;.
\eeq
Here, the first term gets contributions from virtual QCD corrections to all
types of Feynman graphs as in Fig.~\ref{fig:feyn} but the finite second
part originates from virtual QCD corrections to only those Feynman graphs
where two electroweak bosons are attached to the same fermion line, as
shown in Fig.~\ref{fig:feyn}(a). The full expression of this finite part
can be expressed in terms
of the finite parts of the Passarino-Veltman \cite{passarino} $B_0$,
$C_0$ and $D_{ij}$ functions and was given in Eq.~(A1) of 
Ref.~\cite{dieter_z} for the analogous case of $V\to l^+l^-$ decay: 
simply replace one of the two polarization vectors $\epsilon_i^\alpha$ 
of Eq.~(A1) by
the slepton current $\Gamma^\alpha_V$ of Eq.~(\ref{eq:currect}).

\begin{figure}[t]
\centerline{
\epsfxsize=8.0 cm\epsfysize=7.0cm
                     \epsfbox{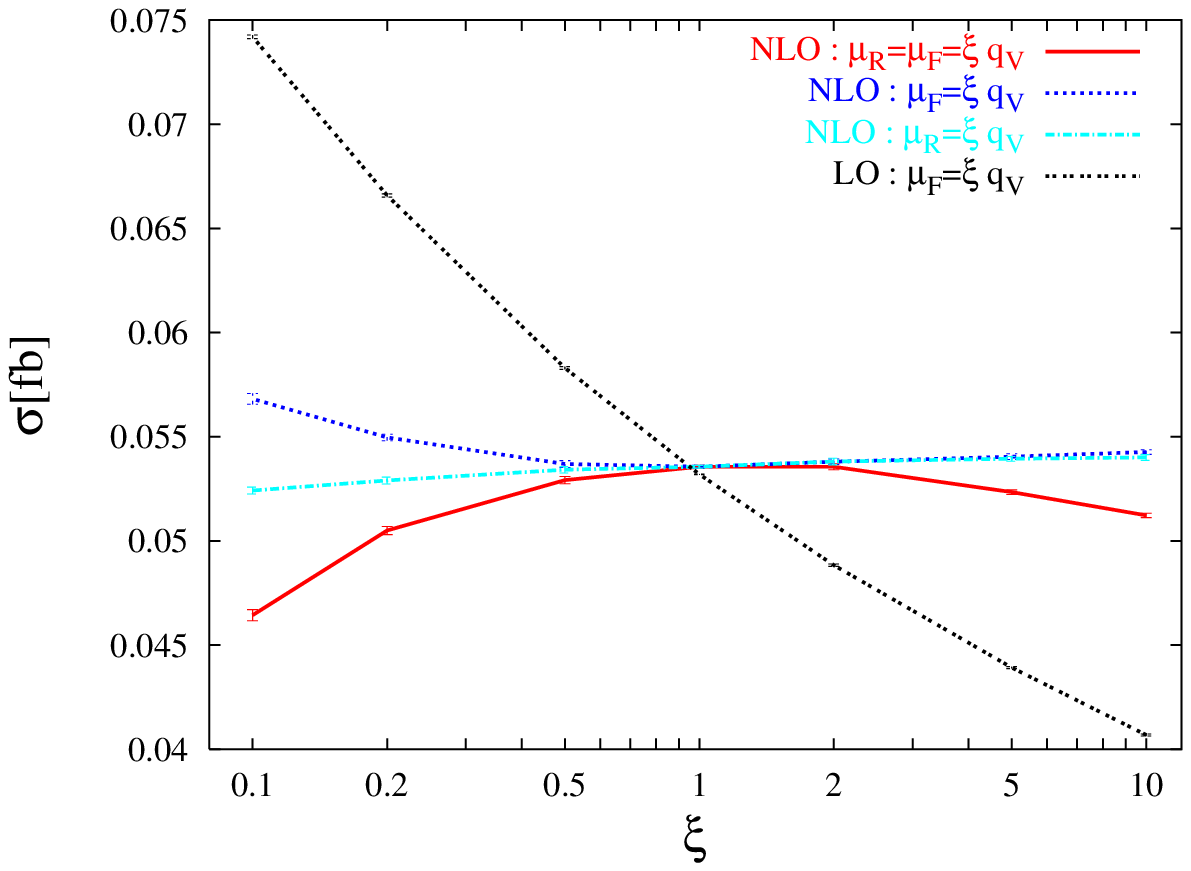}
        \hspace*{-.2cm}
\epsfxsize=8.0 cm\epsfysize=7.0cm
                     \epsfbox{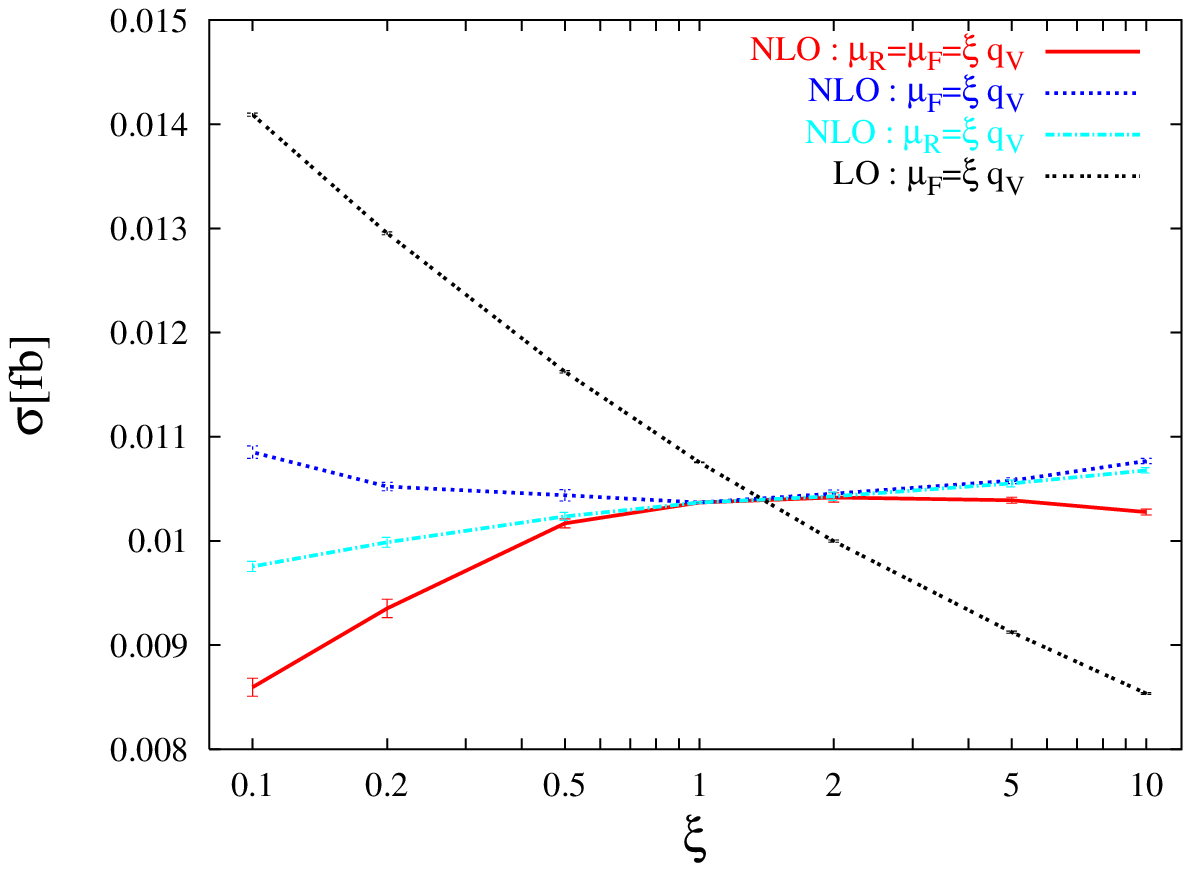}
}

\caption{\it Scale dependence of the $pp \r \slep^+ \slep^- j j X$ 
 cross section at the LHC
 for (a) left-type slepton and (b) right-type slepton at NLO and LO with
 the cuts of Eqns.(\ref{eqn:cut1}, \ref{eqn:cut2}). The
 slepton masses are $m_{\slep_L} = m_{\slep_R}= 200$~GeV.
}
\label{fig:Scale_dep}
\end{figure}

The results obtained for the Born amplitude, the real emission and the
virtual corrections have been tested extensively. For the tree-level
amplitudes (Born and real emission), we have performed a comparison 
to the fully automatically
generated results provided by Smadgraph \cite{rain} and confirmed their
equality numerically. We also checked the invariance of the Born
cross-section under Lorentz transformations. Furthermore, gauge
invariance has been confirmed for the external gluon, within the
numerical accuracy of the program.

\noindent
{\underline{\em Results and Discussions}}:~

The cross-section contributions discussed in the previous section are
implemented in a fully-flexible parton-level Monte Carlo program for EW
$\slep^+ \slep^- j j$ production at NLO QCD accuracy. The program is very
similar to the ones for Hjj, Vjj and VVjj production in VBF described in
Refs.~\cite{dieter_h}, \cite{dieter_z} and \cite{dieter_zz}.  We use the
CTEQ6M parton distributions with $\alpha_s(m_Z) = 0.118$ at NLO, and 
CTEQ6L1 distributions for all LO cross sections. We chose 
$m_Z = 91.188 \GeV$, $m_W = 80.423 \GeV$ and 
$G_F = 1.166 \times 10^{-5} \GeV^{-2}$ as electroweak input
parameters.  Thereof, $\alpha_{QED} = 1/132.54$ and $\sin^2{\theta_W} =
0.22217$ are computed via LO electroweak relations. To reconstruct jets
from final-state partons, the kT algorithm is used with resolution
parameter D = 0.8 \cite{kt_algo}. Throughout, we assume a 
pure $\l \neu$ decay 
of the sleptons, whenever decay distributions are being discussed.

\begin{figure}[t]
\centerline{
\epsfxsize= 8.0 cm\epsfysize=6.0cm
                     \epsfbox{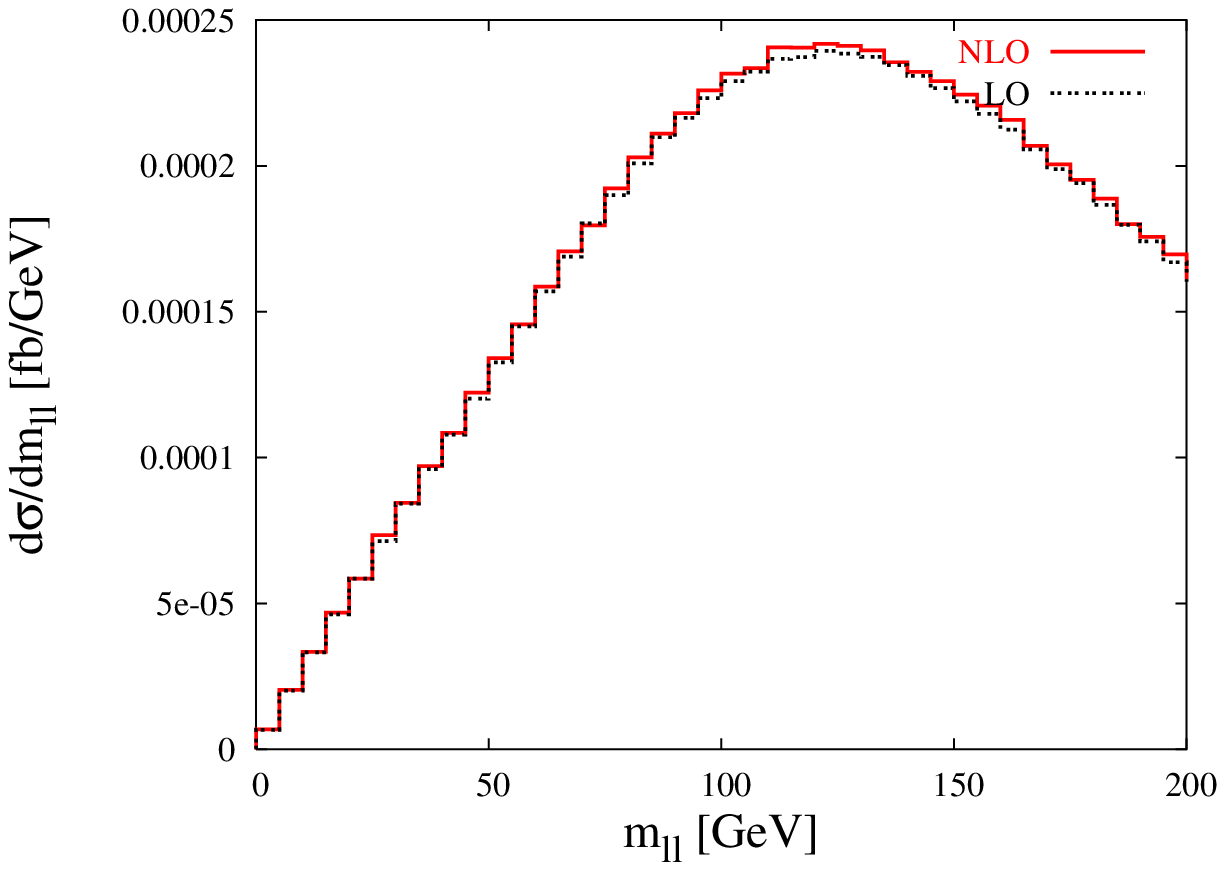}
        \hspace*{-.2cm}
\epsfxsize= 8.0 cm\epsfysize=6.0cm
                     \epsfbox{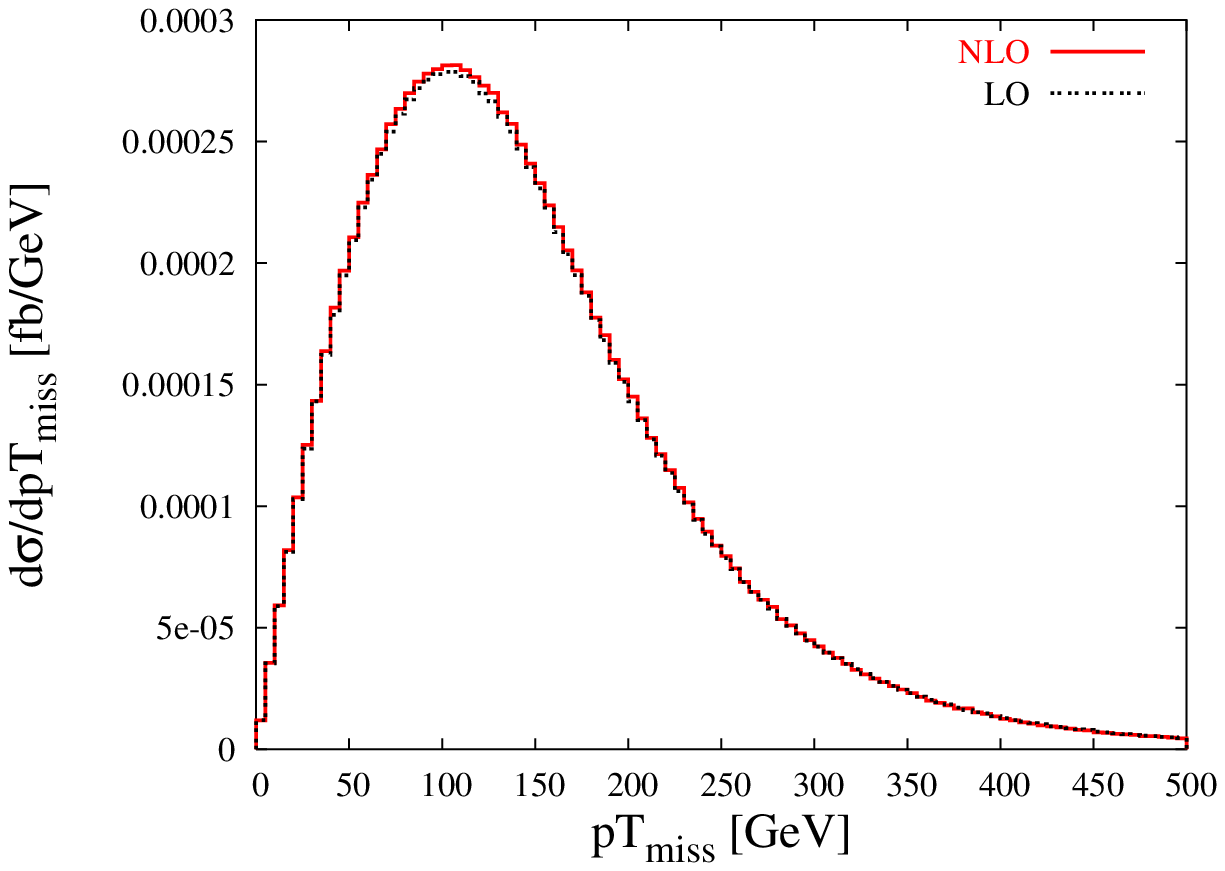}
}

\caption{\it Distributions for (a) daughter lepton invariant mass
  $M_{ll}$ and (b) missing transverse momentum $pT_{miss}$ at NLO
  (solid red) and LO (dashed black). Left type slepton 
  production in the SPS 1a scenario ($\msl = 202 \GeV$) is considered. 
  Renormalization and factorization scales are taken as 
  $\mu_R = \mu_F = q_V$.}
\label{fig:dist1}
\end{figure}
\begin{figure}[ht]
\centerline{
\epsfxsize= 8.0 cm\epsfysize=6.0cm
                     \epsfbox{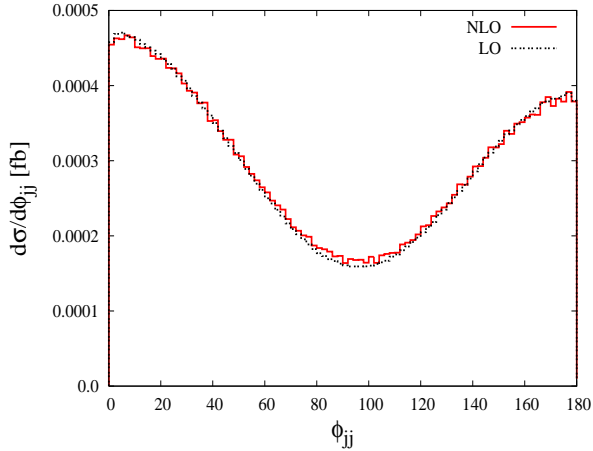}
}

\caption{\it Distributions of the azimuthal angle between the 
  two tagging jets ,  $\phi_{jj}$, at NLO (solid red) and 
  LO (dashed black). Left type slepton
  production in SPS 1a scenario ($\msl = 202 \GeV$) is considered. 
  Renormalization and factorization scales are taken as 
  $\mu_R = \mu_F = q_V$.}
\label{fig:dist2}
\end{figure}

Partonic cross sections are calculated for events with at least two hard
jets, which are required to have
\beq
p_{T_j} \ge 20 \GeV,\qquad |y_j| \le  4.5
\label{eqn:cut1}
\eeq
Here $y_j$ denotes the rapidity of the (massive) jet momentum which is
reconstructed as the four-vector sum of massless partons of
pseudo-rapidity $|\eta| < 5$. These cuts ensure a finite LO differential
cross section for $\slep^+ \slep^- jj$ production, since they enforce finite
scattering angles for the two quark jets. 
The two reconstructed jets of highest
transverse momentum are called 'tagging jets'. At LO, they are the
final-state quarks which are characteristic of vector-boson fusion
processes. Backgrounds to VBF are significantly reduced by requiring a
large rapidity separation of the two tagging jets. We therefore impose
the cut
\beq
\Delta y_{jj} = |y_{j_1} - y_{j_2}| > 4.2
\label{eqn:cut2}
\eeq

Within the above cuts we have calculated the $\slep^+\slep^- jj$ cross
sections at LO and at NLO for the SPS 1a parameter point where slepton
masses are given by $m_{\slep_L}$= 202 GeV, $m_{\slep_R}$= 144
GeV. This point can be parameterized by the mSUGRA model with 
$m_0=100$~GeV, $m_{1/2}=250$~GeV, $A_0 = -100$~GeV, $\tan{\beta}=10$ and 
positive $\mu$~\cite{sps}. We find production cross sections of 
0.0536 (0.0532)~fb for $\slep_L$ production and 0.0242 (0.0249)~fb for 
$\slep_R$ production at NLO (LO) when setting renormalization and 
factorization scales to $\mu_R = \mu_F = q_V$. 
Unfortunately, expected cross 
sections at the LHC are quite small in general, not exceeding 0.1~fb for 
slepton masses above 150~GeV for left-handed sleptons and for
slepton masses above 80~GeV for right-handed sleptons.
In order to compare $\slep_L$ and $\slep_R$ cross sections more directly, 
we have calculated their total production cross sections, within the 
above cuts, for a mass of 200~GeV in both cases.
Fig.~\ref{fig:Scale_dep} illustrates the dependence of these total cross
sections on the renormalization and factorization scales, $\mu_R$
and $\mu_F$, which are taken as multiples of the momentum transfers,
$q_V$, of the $t$-channel electroweak bosons in Fig.~\ref{fig:feyn}, 
$\mu_R = \xi_R q_V$, $\mu_F = \xi_F q_V$.
This choice takes into account that at both LO and NLO the VBF process
can be viewed as a double deep inelastic scattering event, for which the
momentum transfer carried by the exchanged electroweak boson is a
natural scale choice. It leads to K-factors close to unity for
both total cross sections and distributions.

The LO cross section, $\sigma_{LO}$, 
only depend on $\mu_F = \xi_F q_V$.  By varying
the scale factor $\xi_F = \xi$ in the range 0.1 - 10, the value of
$\sigma_{LO}$ changes by around a factor of two, indicating a
substantial theoretical uncertainty of the LO calculation. The strong
scale dependence is reduced at NLO. For $\sigma_{NLO}$, we show three
different cases: $\xi_F = \xi_R = \xi$ (solid red line), $\xi_F = \xi$,
$\xi_R = 1$ (dot-dashed blue line), and $\xi_F = 1$, $\xi_R = \xi$
(dashed green line).  The latter curve illustrates clearly the weak
dependence of $\sigma_{NLO}$ on the renormalization scale, which
can be understood from the fact that $\alpha_s(\mu_R)$ enters only at
NLO.  Also the factorization-scale dependence of the full cross section
is low.  In our following study we fix the scales at $\mu_F = \mu_R =
q_V$, unless noted otherwise.

Two examples for distributions are given in Fig.~\ref{fig:dist1}. 
We show the distributions for (a) the invariant mass, $M_{ll}$, 
of the two charged daughter leptons in the decay $\slep^+\slep^-
\to l^+l^- {p_T}_{miss}$,  and (b) the missing transverse
momentum, ${p_T}_{miss}$.  Results are shown at both LO and NLO and are
virtually indistinguishable with the scale choice $\mu_F = \mu_R = q_V$.
For the illustration in Fig.~\ref{fig:dist1}, left type slepton 
production ($pp \r \sll^+ \sll^- jjX$) in the SPS 1a scenario ($\msl = 202
\GeV$) is considered. 

Within the same set of model parameters, the distribution in the 
azimuthal angle between the two tagging jets, $\phi_{jj}$, is shown
in Fig.~\ref{fig:dist2}. One finds a characteristic dip at $90^0$, a
feature which otherwise is only found in $Hjj$ production when the Higgs
boson couples to gauge bosons via a $HV_{\mu\nu}V^{\mu\nu}$ operator in
the effective Lagrangian~\cite{vera}. $Hjj$ or $Zjj$ production via 
VBF in the SM produces a 
fairly flat $\phi_{jj}$ distribution, while $Hjj$ production via gluon
fusion exhibits a structure very similar to the one shown in
Fig.~\ref{fig:dist2}. Since it has been suggested that 
the $\phi_{jj}$ distribution in VBF events be used in distinguishing 
SM Higgs couplings from anomalous couplings, the possibility that a dip
at $90^0$ might also be produced by the production of two charged
scalars should be kept in mind, should such a feature be discovered at
the LHC.

\begin{figure}[t]
\centerline{
\epsfxsize= 8.0 cm\epsfysize=6.0cm
                     \epsfbox{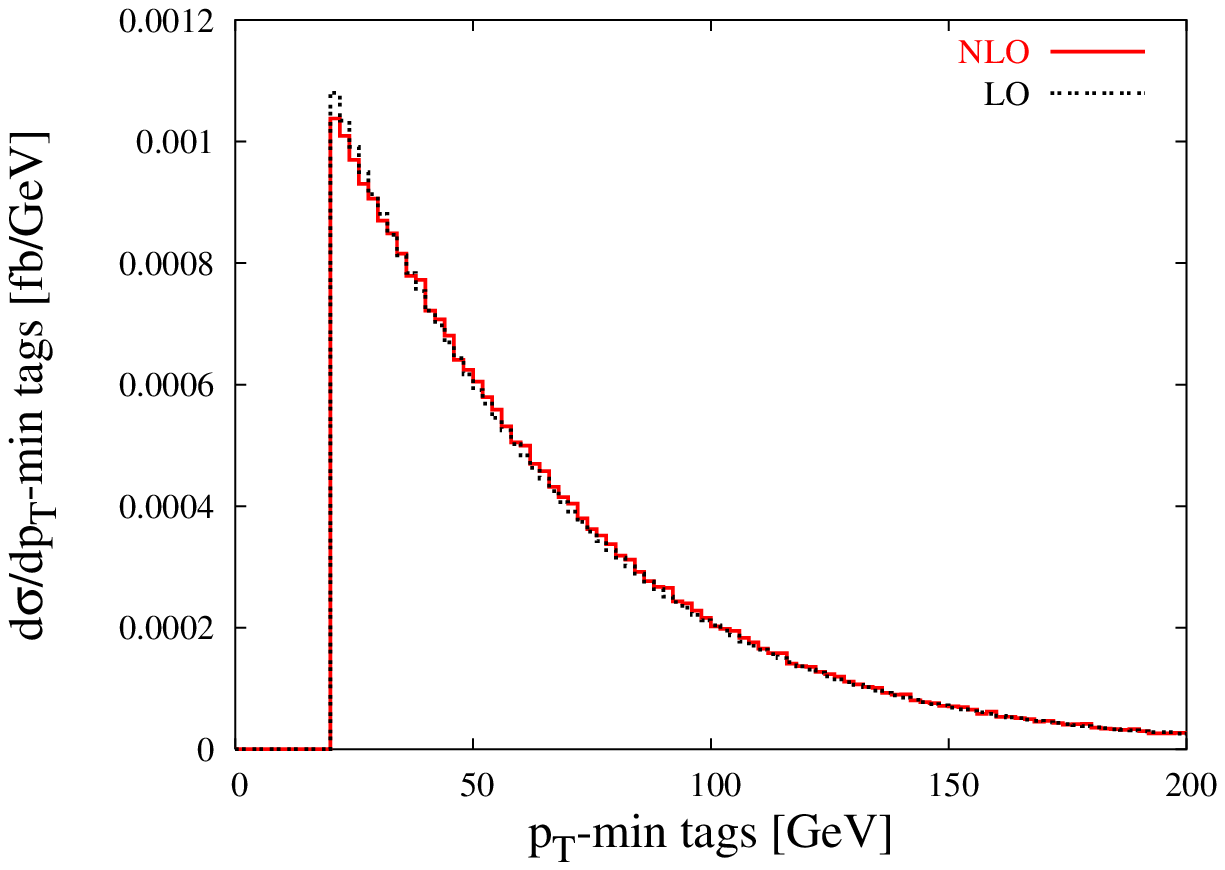}
        \hspace*{-.2cm}
\epsfxsize= 8.0 cm\epsfysize=6.0cm
                     \epsfbox{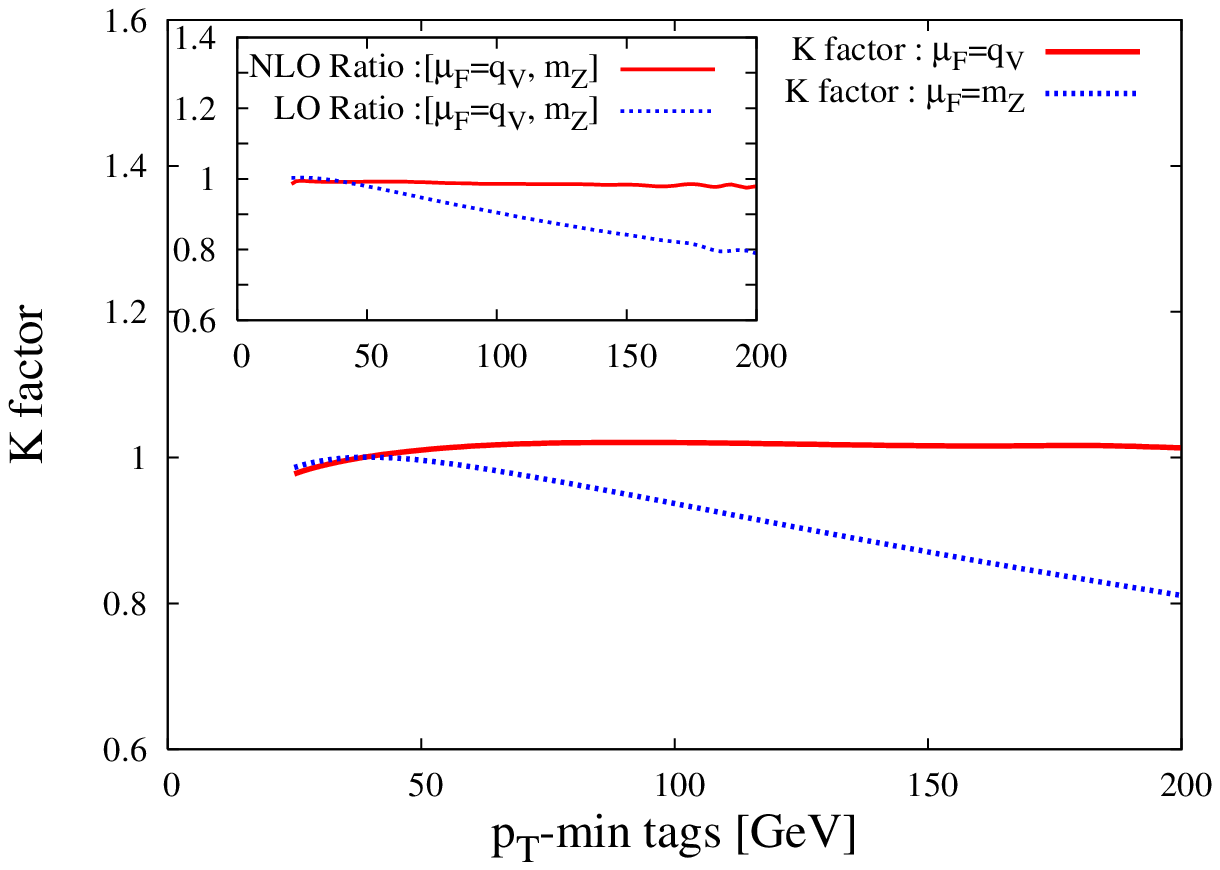}
}

\caption{\it (a) Distribution of the minimum transverse momentum of the
  two tagging jets, $p_{T,min}(tags)$, at NLO (solid red) and LO (dashed
  black) for the scale choice $\mu_R = \mu_F = q_V$. 
  (b) K factor as defined in Eq.~(\ref{eqn:K_fac}) after fixing 
  the factorization scale at the momentum transfer $\mu_F = q_V$
  (solid red) and at $\mu_F = m_Z$ (dashed blue). The inset shows the
  ratio between these two choice of factorization scale at NLO (solid
  red) and at LO (dashed black). Left type slepton production in the SPS
  1a scenario ($\msl = 202 \GeV$) is shown.}
\label{fig:K_fac}
\end{figure}

 A distribution which distinguishes slepton pair production from many 
other VBF processes is the minimum transverse momentum of the two tagging 
jets, $d\sigma/dp_{T,min}(tags)$, which is depicted in 
Fig.~\ref{fig:K_fac}(a). Due to a significant contribution from $t$-channel
photon exchange, this distribution falls quite steeply.
An even steeper fall-off is found for
right type slepton production ($pp \r \slr^+ \slr^- jjX$), 
which can be understood from the fact that $\slr$ has no 
coupling to the $W^{\pm,0}$ eigenstates. 

The shape of the above distribution at LO can differ significantly
from the respective NLO result when scales other than $\mu_R = \mu_F =
q_V$ are used. This is emphasized in Fig.~\ref{fig:K_fac}(b), where we show
the dynamical K factor, defined as
\beq
K(x) = \frac{d\sigma_{NLO}/dx}{d\sigma_{LO}/dx}
\label{eqn:K_fac}
\eeq
for the two choices $\mu_F =q_V$ and $\mu_F =m_Z$ (and $\mu_R=q_V$ 
for the NLO curves in both cases).
While the NLO cross sections differ very little when switching between
the two scale choices (see inset), the effect on the LO cross sections
is quite 
sizable, approaching a 20\% effect at $p_{Tj}\approx 200$~GeV.
This is to illustrate that the choice $\mu_F = q_V$ minimizes the NLO 
corrections in most distributions, by producing a LO prediction which
is close to the true NLO result.

\noindent
{\underline{\em Summary and Conclusions}}:~

In this paper we have presented results for EW slepton pair production 
at NLO QCD
accuracy, obtained with a new parton-level Monte Carlo program. 
The integrated cross sections for this process are consistent with
the results of Ref.~\cite{rain} and show a very moderate K factor. 
While NLO results are
quite stable against scale variations, LO results can change
substantially. We find that the higher order QCD corrections are 
minimized by the scale choice $\mu_F = q_V$ at LO, where $q_V$ is 
the momentum transfer carried by the $t$-channel electroweak bosons.

A second observation concerns the distribution of the azimuthal angle
separation between the two tagging jets in VBF events. The VBF
production of two scalars, as considered here, produces the same type of
dip at $90^0$ as is otherwise observed only for Higgs production with
loop induced couplings to the fusing vector bosons. 

\bigskip
\centerline{\em Acknowledgments}
{\footnotesize P.K. would like to thank B.~J\"ager for many helpful
  discussions in the process of numerical calculation. This research was
  supported in part by the Deutsche Forschungsgemeinschaft under 
  SFB/TR-9 ``Computergest\"utzte Theoretische Teilchenphysik''.}

\bigskip
\bigskip
\bigskip

\end{document}

%% file: diag_generic.tex
{
\unitlength=1.5 pt
\SetScale{1.5}
\SetWidth{0.7}      
\scriptsize    
\hskip 2cm
\begin{picture}(95,99)(0,0)
\Text(15.0,80.0)[r]{$q$}
\ArrowLine(16.0,80.0)(58.0,80.0)
\Text(80.0,90.0)[l]{$q$}
\ArrowLine(58.0,80.0)(79.0,90.0)
\Text(73.0,75.0)[r]{$\gamma,Z$}
\Photon(58.0,80.0)(58.0,60.0){2}{3}
\Text(80.0,70.0)[l]{$\tilde{\ell}^-$}
\DashArrowLine(58.0,60.0)(79.0,70.0){1.0}
\Text(80.0,50.0)[l]{$\tilde{\ell}^+$}
\DashArrowLine(79.0,50.0)(58.0,60.0){1.0}
\Text(15.0,40.0)[r]{$q$}
\ArrowLine(16.0,40.0)(58.0,40.0)
\Text(80.0,30.0)[l]{$q$}
\ArrowLine(58.0,40.0)(79.0,30.0)
\Photon(37.0,80.0)(37.0,40.0){2}{6}
\Text(33.0,60.0)[r]{$W,\gamma,Z$}
\Vertex(58.0,80.0){1.5}
\Vertex(58.0,40.0){1.5}
\Vertex(27.0,80.0){1.5}
\Vertex(27.0,40.0){1.5}
\Text(47,20)[b] {(a)}
\end{picture} \
{} \qquad\allowbreak
\begin{picture}(95,99)(0,0)
\Text(15.0,80.0)[r]{$q$}
\ArrowLine(16.0,80.0)(37.0,80.0)
\Text(80.0,90.0)[l]{$q$}
\ArrowLine(37.0,80.0)(79.0,90.0)
\Text(33.0,70.0)[r]{$W,Z$}
\Photon(37.0,80.0)(37.0,60.0){2}{3}
\Text(50.0,63.0)[b]{$h,H$}
\DashLine(37.0,60.0)(58.0,60.0){3.0}
\Text(80.0,70.0)[l]{$\tilde{\ell}^-$}
\DashArrowLine(58.0,60.0)(79.0,70.0){1.0}
\Text(80.0,50.0)[l]{$\tilde{\ell}^+$}
\DashArrowLine(79.0,50.0)(58.0,60.0){1.0}
\Text(33.0,50.0)[r]{$W,Z$}
\Photon(37.0,60.0)(37.0,40.0){2}{3}
\Text(15.0,40.0)[r]{$q$}
\ArrowLine(16.0,40.0)(37.0,40.0)
\ArrowLine(37.0,40.0)(79.0,30.0)
\Text(80.0,30.0)[l]{$q$}
\Text(47,20)[b] {(b)}
\end{picture} \\ 
\begin{picture}(95,99)(0,0)
\Text(15.0,90.0)[r]{$q$}
\ArrowLine(16.0,90.0)(58.0,90.0)
\Text(80.0,90.0)[l]{$q$}
\ArrowLine(58.0,90.0)(79.0,90.0)
\Text(53.0,80.0)[r]{$W,\gamma,Z$}
\Photon(58.0,90.0)(58.0,70.0){2}{3}
\Text(80.0,70.0)[l]{$\tilde{\ell}^-$}
\DashArrowLine(58.0,70.0)(79.0,70.0){1.0}
\Text(54.0,60.0)[r]{$$}
\DashArrowLine(58.0,50.0)(58.0,70.0){1.0}
\Text(80.0,50.0)[l]{$\tilde{\ell}^+$}
\DashArrowLine(79.0,50.0)(58.0,50.0){1.0}
\Text(53.0,40.0)[r]{$W,\gamma,Z$}
\Photon(58.0,50.0)(58.0,30.0){2}{3}
\Text(15.0,30.0)[r]{$q$}
\ArrowLine(16.0,30.0)(58.0,30.0)
\Text(80.0,30.0)[l]{$q$}
\ArrowLine(58.0,30.0)(79.0,30.0)
\Text(47,20)[b] {(c)}
\end{picture} \ 
\begin{picture}(95,99)(0,0)
\Text(15.0,80.0)[r]{$q$}
\ArrowLine(16.0,80.0)(58.0,80.0)
\Text(80.0,90.0)[l]{$q$}
\ArrowLine(58.0,80.0)(79.0,90.0)
\Text(53.0,70.0)[r]{$W,\gamma,Z$}
\Photon(58.0,80.0)(58.0,60.0){2}{3}
\Text(80.0,70.0)[l]{$\tilde{\ell}^-$}
\DashArrowLine(58.0,60.0)(79.0,70.0){1.0}
\Text(80.0,50.0)[l]{$\tilde{\ell}^+$}
\DashArrowLine(79.0,50.0)(58.0,60.0){1.0}
\Text(53.0,50.0)[r]{$W,\gamma,Z$}
\Photon(58.0,60.0)(58.0,40.0){2}{3}
\Text(15.0,40.0)[r]{$q$}
\ArrowLine(16.0,40.0)(58.0,40.0)
\Text(80.0,30.0)[l]{$q$}
\ArrowLine(58.0,40.0)(79.0,30.0)
\Text(47,20)[b] {(d)}
\end{picture} \
\begin{picture}(95,99)(0,0)
\Text(15.0,80.0)[r]{$q$}
\ArrowLine(16.0,80.0)(37.0,80.0)
\Text(80.0,90.0)[l]{$q$}
\ArrowLine(37.0,80.0)(79.0,90.0)
\Text(33.0,70.0)[r]{$W$}
\Photon(37.0,80.0)(37.0,60.0){2}{3}
\Text(49.0,65.0)[b]{$\gamma,Z$}
\Photon(37.0,60.0)(58.0,60.0){2}{3}
\Text(80.0,70.0)[l]{$\tilde{\ell}^-$}
\DashArrowLine(58.0,60.0)(79.0,70.0){1.0}
\Text(80.0,50.0)[l]{$\tilde{\ell}^+$}
\DashArrowLine(79.0,50.0)(58.0,60.0){1.0}
\Text(33.0,50.0)[r]{$W$}
\Photon(37.0,60.0)(37.0,40.0){2}{3}
\Text(15.0,40.0)[r]{$q$}
\ArrowLine(16.0,40.0)(37.0,40.0)
\Text(80.0,30.0)[l]{$q$}
\ArrowLine(37.0,40.0)(79.0,30.0)
\Text(47,20)[b] {(e)}
\end{picture} \ 
}